\documentclass[10pt]{emulateapj}
\usepackage{apjfonts}
\def\Meszaros{M\'esz\'aros~}

\begin{document}

\title{GRB 080916C:  On the radiation origin of the prompt emission from KeV/MeV to  GeV}

\author{Xiang-Yu  Wang\altaffilmark{1},  Zhuo Li\altaffilmark{2,3}, Zi-Gao Dai\altaffilmark{1} and Peter M\'esz\'aros\altaffilmark{4,5}}
\altaffiltext{1}{Department of Astronomy, Nanjing University,
Nanjing 210093, China} \altaffiltext{2}{Department of Astronomy,
Peking University, Beijing 100871, China} \altaffiltext{3} {Kavli
Institute for Astronomy and Astrophysics, Peking University,
Beijing 100871, China} \altaffiltext{4}{Department of Astronomy
and Astrophysics, Pennsylvania State University, University Park,
PA 16802, USA}\altaffiltext{5}{Department of Physics, Pennsylvania
State University, University Park, PA 16802, USA}

\begin{abstract}
Fermi observations of high-energy gamma-ray emission from GRB
080916C shows that its spectrum is consistent with the Band
function from MeV to tens of GeV. Assuming one single emission
mechanism dominates in the whole energy range, we show that this
spectrum is consistent with synchrotron origin by
shock-accelerated electrons. The simple electron inverse-Compton
model and the hadronic model are found to be less viable. In the
synchrotron scenario, the synchrotron self-Compton scattering is
likely to be in the Klein-Nishina regime and therefore the
resulting high-energy emission is subdominant, even though the
magnetic field energy density is lower than that in relativistic
electrons. The Klein-Nishina inverse-Compton cooling may also
affect the low-energy electron number distribution and hence
results in a low-energy synchrotron photon spectrum
$n(\nu)\propto\nu^{-1}$ below the peak energy. Under the framework
of the electron synchrotron interpretation, we constrain the shock
microphysical parameters and derive a lower limit of the upstream
magnetic fields. The detection of synchrotron emission extending
to about 70 GeV in the source frame in GRB 080916C favors the Bohm
diffusive shock acceleration if the bulk Lorentz factor of the
relativistic outflow is not significantly greater than thousands.

\end{abstract}

\keywords{gamma rays: bursts}

\section{Introduction}
It was recently reported that Fermi satellite has detected strong
$>100{\rm MeV}$ emission from a very energetic long-duration burst
GRB080916C (Abdo et al. 2009).  At a redshift of $z=4.35\pm0.15$
(Grenier et al. 2009), the burst is the most energetic one ever,
with an isotropic gamma-ray energy $E_\gamma\simeq 8\times10^{54}
{\rm ergs}$, which is released over a duration of $T_{90}\simeq 60
{\rm s}$. Equally remarkably, more than ten photons with energy
above GeV are detected, with the highest energy one at 13 GeV (in
the observer frame). The  spectra of all five time intervals,
designated as times a-e in the light curves of GRB080916C (Abdo et
al. 2009), are well fit by the empirical Band function (Band et
al. 1993), which smoothly joins low- and high-energy power laws.
The high-energy power law extends to GeV energies, without any
additional spectral component visible. The peak energy of the
spectra during these intervals is around
$\varepsilon_p\simeq400{\rm KeV}-1{\rm MeV}$. Except during the
first time interval, the low-energy and high-energy photon
spectral indices of the prompt emission are constant and
consistent with $\alpha\simeq-1.0$ and $\beta\simeq-2.2$
respectively. With its high temporal resolution, INTEGRAL detected
the temporal variability of the KeV/MeV emission on time scales as
short as 100 ms with high statistical significance (Greiner et al.
2009). So the variability timescale in the local source frame is
$t_v\la100{\rm ms}/(1+z)=20{\rm ms}$.

The nonthermal synchrotron radiation by electrons has been
suggested to be a possible mechanism for the 10 KeV-MeV emission
(see \Meszaros 2006 and Zhang 2007 for recent reviews), but one
famous problem remains so far, i.e. the low-energy photon spectral
index $\alpha$ is incompatible with the index $-3/2$ that is
expected from fast-cooling electrons (e.g. Preece et al. 1996;
Ghisellini et al. 2000). Electron inverse Compton emission has
been a competitive mechanism (e.g. Panaitescu \& \Meszaros 2000).
The fact that one single spectral component fits the spectrum of
the prompt emission from 10 KeV to GeV in GRB080916C suggests that
one emission mechanism dominates in this whole energy
range{\footnote{Dropping the assumption of the same origin for
both MeV and high energy emission, Li (2008) explained the high
energy emission in GRB080916C as delayed, ``residual" emission
from subsequent collisions at larger and larger radii in the
baryonic outflow.}}. In this {\it Letter}, we study the constraint
that this puts on the emission mechanism. Abdo et al. (2009)
mentioned as one of the possibilities that the delay of
high-energy gamma-ray emission relative to low-energy emission in
GRB080916C could be a result of longer acceleration time needed
for higher energy protons or nuclei in hadronic emission models.
In accordance with this, we also study whether such hadronic
models could be a possible mechanism that produces the KeV/MeV to
GeV emission in GRB080916C.

\section{The  synchrotron model and parameter constraints}
Assuming that in GRB shocks, fractions of $\epsilon_B$ and
$\epsilon_e$ of the shock internal energy are converted into the
energy in the magnetic field and  electrons, respectively. To
ensure a high radiation efficiency for the prompt emission, it is
usually assumed that the electrons are rapidly cooling, so the
energy density in gamma-ray emission $U_\gamma$ is equal to the
electron energy density $U_e$. The magnetic field is given by
\begin{equation}
\frac{B^2}{8\pi}=\left(\frac{\epsilon_B}{\epsilon_e}\right)U_\gamma=\left(\frac{\epsilon_B}{\epsilon_e}\right)\frac{L_\gamma}{4\pi
R^2 c \Gamma^2},
\end{equation}
where $R$ is the radius of the shock, $L_\gamma$ is the luminosity
in gamma-ray emission and $\Gamma$ is the bulk Lorentz factor. The
detection of GeV photons suggest that the emission region has a
bulk Loentz factor $\Gamma\ga 10^3$(Grenier et al. 2009; Abdo et
al. 2009) {\footnote{The limit on the bulk Lorentz factor would be
more stringent when we have an IC TeV component whose flux is
above the synchrotron extension, as is the case in the synchrotron
scenario. Using the formula in Lithwick \& Sari (2001), however,
one can find that the limit is only increased by a factor smaller
than two.}}. From the casuality constraint, the emission radius is
$R=\Gamma^2 c t_v$. In the synchrotron model for the 10 KeV-GeV
emission, by use of $\varepsilon_p=h\nu_{
p}=\phi_{\nu}\frac{3hqB}{4\pi m_e c}\gamma_m^2\Gamma$,  one can
derive the Lorentz factor of electrons that radiate at the GRB
peak energy $\varepsilon_p$,
\begin{equation}
\begin{array}{ll}
\gamma_m=(\frac{4\pi m_e c\varepsilon_p}{3\phi_\nu hq})^{1/2}
(\frac{\epsilon_e}{\epsilon_B})^{1/4}(\frac{2 L_\gamma}{R^2
c})^{-1/4} \varepsilon_p^{1/2}\\
=2.5\times10^3 (\frac{\epsilon_e}{\epsilon_B})^{1/4} L_{\gamma,
54}^{-1/4} \Gamma_3 t_{v,-2}^{1/2}(\frac{\varepsilon_p}{2 {\rm
MeV}})^{1/2},
\end{array}
\end{equation}
where $\phi_\nu\simeq0.5$ is the coefficient defined in Wijers \&
Galama (1999) and $q$ is the electron charge. Define $\gamma_{\rm
T}$ as the Lorentz factor of electrons below which the scattering
with peak energy photons are in the Thomson scattering regime,
i.e.
\begin{equation}
\gamma_{\rm T}=\Gamma m_e c^2/ \varepsilon_p=250 \Gamma_3
(\varepsilon_p/2 {\rm MeV})^{-1}.
\end{equation}
Unless $\epsilon_e\la10^{-4}\epsilon_B$, which is unreasonable in
terms of the burst energetics, the IC scattering between
$\gamma_m$ electrons and the bulk of the gamma-ray emission should
be in the Klein-Nishina (KN) regime.

The KN Compton cooling of electrons may affect the low-energy
electron distribution at $\gamma_{\rm T}<\gamma<\gamma_m$ and
hence affect the low-energy spectral slope below $\varepsilon_p$
(e.g. Rees 1967; Derishev et al. 2003), as we show below. Consider
a population of electrons injected into a box with magnetic field
$B$ in a power law form $N(\gamma)\propto \gamma^{-p}$ for
$\gamma\ge\gamma_m$. These electrons will cool down rapidly
through synchrotron and/or IC radiation{\footnote{ The synchrotron
cooling time of $\gamma_m$ electrons is shorter than the dynamic
time $t'_d=R/\Gamma c=\Gamma t_v$ as long as
${\epsilon_e}/{\epsilon_B}\la 10^{6}
\Gamma_3^{-16/3}t_{v,-2}^{-2}({\varepsilon_p}/{2 {\rm
MeV}})^{2/3}L_{\gamma,54}$. }}. As the electron cools, its energy
changes as $\gamma$, so we have
\begin{equation}
\nu F_\nu \left[1+k(\gamma){U_\gamma}/{U_B}\right] \propto \gamma
\end{equation}
for $\gamma<\gamma_m$, where $\nu$ is the synchrotron frequency of
$\gamma$-electrons and $k(\gamma)$ accounts for the reduction of
the effective photon energy density for IC scattering of the
$\gamma$-electrons due to the KN effect. Define $h\nu_{KN}=\Gamma
m_e c^2/\gamma$ as the critical energy of the photons with which
the IC scattering of $\gamma$-electrons is just in the KN regime.
For a low-energy photon spectrum of the form $ \nu F_\nu\propto
\nu^\delta$ ($\nu<\nu_p$), we have
\begin{equation}
k(\gamma)\simeq
\frac{U_\gamma(\nu<\nu_{KN})}{U_\gamma}=(\frac{\nu_{KN}}{\nu_p})^{\delta}=(\frac{\gamma}{\gamma_{\rm
T}})^{-\delta}
\end{equation}
for $\gamma_{\rm T}<\gamma<\gamma_{m}$ and $k\simeq 1$ for
$\gamma<\gamma_{\rm T}$. In the case of $\gamma_{\rm
T}<\gamma<\gamma_m$, if $k U_\gamma/U_B\gg 1 $, i.e. the IC
cooling is still dominant even though the scatterings are in the
KN regime, one can obtain $\nu F_\nu \propto
\gamma^{\delta+1}\propto \nu^{(\delta+1)/2}$, where we have used
$\nu\propto \gamma^2$ in the last step. Equating this synchrotron
spectral index  with the initially assumed spectral index
$F_\nu\propto \nu^{\delta-1}$, one can derive
\begin{equation}
\delta=1,  F_\nu\propto \nu^0.
\end{equation}
This can explain the observed low-energy photon index of
$\alpha=-1.0$ in GRB080916C if the condition,  $U_B\la k(\gamma)
U_\gamma$ for $\gamma_{\rm T}<\gamma<\gamma_m$, is satisfied.

On the other hand,  the high-energy spectrum of GRB080916C above
$\varepsilon_p$  is consistent with the synchrotron spectrum
produced by fast-cooling electrons above $\gamma_m$, i.e.
$F_\nu\propto \nu^{-p/2}$ with $p=-2(1+\beta_2)=2.4$. The
dominance of  synchrotron cooling  above $\varepsilon_p$ implies
that  $U_B\ga k(\gamma) U_\gamma$ for electrons with Lorentz
factor $\gamma\ga \gamma_m$. Therefore, we find that, at
$\gamma=\gamma_m$, $U_B\simeq k U_\gamma$. Since $U_\gamma\simeq
U_e$  for fast-cooling electrons,   the requirement $U_B\simeq
k(\gamma_m) U_\gamma$ translates into $\epsilon_e/\epsilon_B\simeq
\gamma_m \varepsilon_p/\Gamma m_e c^2$, which gives
\begin{equation}
\frac{\epsilon_e}{\epsilon_B}\simeq 20 L_{\gamma, 54}^{-1/3}
 t_{v,-2}^{2/3}(\frac{\varepsilon_p}{2 {\rm MeV}})^{2} .
\end{equation}
{Note that the transition region between the two asymptotic
power-laws at low and high energy ends in the Band function is
rather wide, so the above  requirement, $U_B\simeq k(\gamma_m)
U_\gamma$, should be regarded as an order of magnitude of
estimate. In addition, this requirement applies only to large
$\varepsilon_p$ bursts, because for low $\varepsilon_p$ bursts,
the IC scattering may be no longer in the Kelin-Nishina regime. A
signature that high $\varepsilon_p$ bursts have $\alpha$
preferentially close to $-1$ can be seen in the analysis of Preece
et al. (1996). For some low $\varepsilon_p$ bursts that have
$\alpha\simeq-1$, some other mechanisms may be at work.}

For electrons with $\gamma\la \gamma_{\rm T}$, the IC scatterings
with peak energy photons are in the Thomson scattering regime, so
$k(\gamma)=1$ and $N(\gamma)\propto \gamma^{-2}$, leading to a
conventional fast-cooling photon spectrum of $F_{\nu}\propto
\nu^{-1/2}$. Observations show a single power law spectrum
$F_{\nu}\propto \nu^{0}$ from $10$ keV to $\sim {\rm MeV}$ in
GRB080916C, implying that $\gamma_m/\gamma_{\rm T}\ga10$, and one
can therefore obtain  a constraint
\begin{equation}
\frac{\epsilon_e}{\epsilon_B}\ga 1 L_{\gamma,54}
t_{v,-2}^{-2}(\frac{\varepsilon_p}{2 {\rm MeV}})^{-6}
\end{equation}

Due to that the IC scatterings between $\gamma_m$ electrons and
the peak energy photons with energy $\varepsilon_p$ are in the KN
regime, the IC emission peaks at
\begin{equation}
\begin{array}{ll}
h\nu^{IC}_{p}=\Gamma \gamma_m m_e c^2
\\=1(\frac{\epsilon_e}{\epsilon_B})^{1/4} L_{\gamma, 54}^{-1/4}
\Gamma_3^2  t_{v,-2}^{1/2}(\frac{\varepsilon_p}{2 {\rm
MeV}})^{1/2} {\rm TeV},
\end{array}
\end{equation}
with a flux  $\nu F_\nu^{IC}
(\varepsilon_\gamma=h\nu^{IC}_{p})=Y(\gamma_m)\nu_p
F_{\nu_p}\simeq\nu_p F_{\nu_p}$, where $Y$ is the Compton
parameter. For a flat synchrotron spectrum with $\beta\simeq -2$
above $\nu_p$, it is natural that the IC component is not seen at
high energies since $\nu F^{syn}_{\nu}(\varepsilon_h=70 {\rm
GeV})\ga \nu F_\nu^{IC}(\varepsilon_h=70 {\rm GeV})$ for
GRB080916C.

In the above, we have not assumed any model for the dissipation
mechanism of the shocks. In the popular internal shock scenario,
the typical Lorentz factors of the shocked electrons is
$\gamma_m=\epsilon_e (m_p/m_e) \Gamma_{sh}$, where $\Gamma_{sh}$
is the shock  Lorentz factor, which is equal to the relative
Lorentz factor of the two colliding shells. For GRB080916C, we
have obtained a constraint $\gamma_m\simeq 5\times10^3 \Gamma_3
L_{\gamma, 54}^{-1/3} t_{v,-2}^{2/3}(\frac{\varepsilon_p}{2 {\rm
MeV}})$. So if internal shock applies to GRB080916C, we would need
a large relative  Lorentz factor, $\Gamma_{sh}\simeq 8
(3\epsilon_{e})^{-1}\Gamma_3 L_{\gamma, 54}^{-1/3}
t_{v,-2}^{2/3}(\frac{\varepsilon_p}{2 {\rm MeV}})$. This could be
caused by the interaction among the shells whose Lorentz factors
have a large contrast (Yu et al. 2009).  Of course, the shock
could also arise from the magnetic reconnection or turbulence
(e.g. Thompson 1994; \Meszaros \& Rees 1997; Lyutikov \& Blandford
2003; Narayan \& Kumar 2008; Zhang \& Pe'er 2009) and we do not
have the estimate of the shock Lorentz factor from the first
principle.

\section{Alternative models for MeV-10 GeV emission?}
\subsection{The one-zone SSC  scenario}
Let's explore whether the simple IC scenario (i.e. one zone
synchrotron self-Compton (SSC) scenario) can explain the single
power-law spectrum from MeV to 10 GeV in GRB080916C. Suppose that
the first-order SSC of electrons with energy $\gamma_m$ produce
the peak emission $\varepsilon_p=2{\rm MeV}$. Since this IC
emission is not hidden by the synchrotron emission, one will
expect that the 2nd order IC emission appears at high-energy if
the 2nd-order IC peak is located within the observation energy
window and that the 2nd-order IC scattering is still in the
Thomson scattering regime. The fact that we did not see the
2nd-order IC component implies that $\gamma_m^2 \varepsilon_p\ga70
{\rm GeV}$ or $\gamma_m \varepsilon_p\ga \Gamma m_e c^2$, so we
have $\gamma_m\ga190 (\varepsilon_p/2{\rm MeV})^{-1/2}$ or
$\gamma_m\gg\gamma_{\rm T}=250 \Gamma_3 (\varepsilon_p/2{\rm
MeV})^{-1}$. Since $\varepsilon_p=h\nu_{syn,p}\gamma_m^2$, one
obtains the synchrotron peak frequency at $h\nu_{syn,p}=
55(\gamma_m/190)^{-2}(\varepsilon_p/2{\rm MeV}){\rm eV}$. Then one
can obtain an upper limit of the magnetic filed $B=h\nu_{syn,
p}/(\phi_{\nu}\frac{3q}{4\pi m_e c}\gamma_m^2\Gamma)= 140
(\gamma_m/190)^{-4}\Gamma_3^{-1}(\varepsilon_p/2{\rm MeV}) {\rm
G}$. With this magnetic field, we can derive an upper limit for
$\epsilon_B/\epsilon_e$. In the case that 2nd-order IC is still in
the Thomson regime, $U_e= (Y+1)U_\gamma$, so from
$\epsilon_B/\epsilon_e=U_B/U_e\simeq U_B/(Y U_\gamma)$, we obtain
$Y(\epsilon_B/\epsilon_e)=3\times10^{-5}(\gamma_m/190)^{-8}(\varepsilon_p/2{\rm
MeV})^2 L_{\gamma, 54}^{-1}\Gamma_3^{4}t_{v,-2}^2$. By use of
$Y=(\epsilon_e/\epsilon_B)^{1/3}$ (Kobayashi et al. 2007), one get
$Y= 170 (\gamma_m/190)^{4}(\varepsilon_p/2{\rm
MeV})^{-1}L_{\gamma, 54}^{1/2}\Gamma_3^{-2}t_{v,-2}^{-1}$. As
$\gamma_m\ga 190$, so the radiation energy in the 2nd-order IC
will be $E_{2nd, IC}=Y E_\gamma\ga 1.3\times10^{57}{\rm erg}$,
which is too large to be realistic. Such an energy crisis problem
has also been found in the case of GRB080319B for the IC scenario
of the prompt MeV emission (Piran et al. 2008).

On the other hand, if the 2nd-order IC is already  in the deep KN
regime (for $\gamma_m\gg \gamma_{T}=250 \Gamma_3$),
$\epsilon_B/\epsilon_e=U_B/U_e=U_B/U_\gamma$.  The 2nd-order
Compton $Y$ parameter is $Y_{\rm
2nd}=(\epsilon_e/\epsilon_B)^{1/2}(\gamma_m/\gamma_{\rm
T})^{-\delta}$ for a spectrum $\nu F_\nu\propto \nu^{\delta}$
below $\varepsilon_p$ (see Eq.5). To get $Y_{\rm 2nd}\la 1$, one
needs $\gamma_m\ga (\epsilon_e/\epsilon_B)^{1/2} \gamma_{\rm
T}=250 (\epsilon_e/\epsilon_B)^{1/2}\Gamma_3(\varepsilon_p/2{\rm
MeV})^{-1}$ for $\delta=1$. So from $h\nu_{syn,p}=
55(\gamma_m/190)^{-2}(\varepsilon_p/2{\rm MeV}){\rm eV}$, one can
obtain $B\la 45
(\epsilon_e/\epsilon_B)^{-2}\Gamma_3^{-1}(\varepsilon_p/2{\rm
MeV}) {\rm G}$. Combing this upper limit  with the equipartition
assumption in Eq.(1), one can get $\epsilon_e/\epsilon_B\la 0.02
\Gamma_3^{4/3} t_{v,-2}^{2/3} (\varepsilon_p/2{\rm
MeV})^{2/3}L_{\gamma,54}^{-1/2}$, which is in conflict with the
precondition $\epsilon_e/\epsilon_B=Y^2\ga 1$. This means that
significant suppression of the 2nd-order IC emission by KN
scattering can not be fulfilled. So we conclude that the the
simple one-zone SSC model does not work for the MeV to 10 GeV
emission in GRB080916C.

\subsection{The hadronic scenario}
We first study whether the proton synchrotron emission can produce
the MeV-10 GeV emission of GRB080916C. The photon spectrum index
above $\varepsilon_p$, $\beta=-2.2$, implies that  the proton
distribution index is $p\simeq2.4$ for fast-cooling protons or a
very steep index $p\simeq 3.4$ for slow-cooling ones. In the
proton synchrotron scenario, the observed peak emission at
$\varepsilon_p$ is produced by protons with a Lorentz factor of
$\gamma_p=(\frac{4\pi \varepsilon_p m_p c}{3 \phi_{\nu} q h B
\Gamma})^{1/2}$. The synchrotron cooling time of these protons in
the comoving frame is $t'_{syn}={6\pi m_p^3 c}/({\sigma_T m_e^2
\gamma_p B^2})$. Define that the magnetic field energy density is
a fraction of $\xi_B$ of the comoving frame photon energy density,
i.e. $U_B=\xi_B U_\gamma$. So the synchrotron radiation efficiency
of the $\gamma_p$ protons is $\eta=\min[1, {t'_d}/{t'_{syn}}]$,
where $t'_d=R/\Gamma c$ is the dynamic time in the comoving frame,
which  is equal to the comoving frame variability time,
$t'_d=t'_v=\Gamma t_v$. As long as ${t'_v}/{t'_{syn}}\la 1$, we
have a radiation efficiency for $\gamma_p$ protons
\begin{equation}
\eta(\gamma_p)=\frac{t'_v}{t'_{syn}}=3\times 10^{-4} \xi_B^{3/4}
L_{\gamma,54}^{3/4}t_{v,-2}^{-1/2}\Gamma_3^{-4}(\frac{\varepsilon_p}{2
{\rm MeV}})^{1/2}
\end{equation}
Such a low radiation efficiency implies an unrealistically large
energy in protons, a factor of $1/\eta\simeq
3\times10^3\xi_B^{-3/4}$ higher than the energy in gamma-rays. The
radiation efficiency is quite low ($\eta\sim 2\times10^{-3}$) even
for protons that produce the high-energy gamma-rays of energy $\ga
100{\rm MeV}$. So there is no room for the proton synchrotron
model even in the assumption that the high-energy gamma-ray
emission belongs to a different component than the MeV component.
If the spectrum above $\varepsilon_p$ is interpreted as arising
from fast-cooling protons, as required in the case of $2\la p\la
3$, one would need $\xi_B=U_B/U_\gamma \ga 5\times10^4
L_{\gamma,54}^{-1}t_{v,-2}^{2/3}\Gamma_3^{16/3}(\frac{\varepsilon_p}{2
{\rm MeV}})^{-2/3}$, which is also unreasonable, as the total
energy in the magnetic field is too large to be realistic for a
GRB.

Let's also explore the scenario of the secondary emission from
hadronic photopion process. Detection of high-energy gamma-rays of
energy greater than 10 GeV puts a constraint on the opacity of
$\gamma\gamma$ absorption. As the hadronic $p\gamma$ opacity is
related with the $\gamma\gamma$ opacity, a higher maximum photon
energy, hence a lower $\gamma\gamma$ absorption opacity, would
imply a lower hadronic radiation efficiency (e.g. Dermer et al.
2008). It is useful to express the hadronic $p\gamma$ efficiency
as a function of the the pair production optical depth
$\tau_{\gamma\gamma}$. Following Waxman \& Bahcall (1997), the
optical depth for pair production of a photon of energy
$\varepsilon_h$ is
$\tau_{\gamma\gamma}(\varepsilon_h)=\frac{R}{\Gamma
l_{\gamma\gamma}}=\frac{R}{ \Gamma}\frac{\sigma_T}{16}
\frac{U_\gamma \varepsilon_h}{\Gamma (m_e c^2)^2}$, where
$l_{\gamma\gamma}$ is the mean free path. For the simplicity of
calculation, we have assumed a photon spectrum $\beta_2=-2$ above
$\varepsilon_p$, which is a good approximation for GRB080916C. The
fraction of energy lost by protons to pions is
$f_{\pi}\simeq\frac{R}{\Gamma}\frac{U_\gamma}{2\varepsilon'_p}
 \sigma_{p\gamma} \xi_{\rm peak}$ for protons with energy greater than $E_p=6\times
10^{16}\Gamma_3^2 (\varepsilon_p/2{\rm MeV})^{-1} {\rm eV}$
(Waxman \& Bahcall 1997), where $\sigma_{p\gamma}\simeq 5\times
10^{-28} {\rm cm^2}$ is the cross section of the $p\gamma$
reaction at the $\Delta$ resonance and $\xi_{\rm peak}\simeq 0.2$
is the fraction of proton energy loss in one interaction. So the
maximum photopion efficiency is
\begin{equation}
\begin{array}{ll}
f_\pi=2\times10^{-3}\Gamma_3^2(\frac{\varepsilon_h}{70 {\rm
GeV}})^{-1} (\frac{\varepsilon_p}{\rm 2
MeV})^{-1}\tau_{\gamma\gamma}(\varepsilon_h) \\
=2\times10^{-3} ({\Gamma}/{\Gamma_{\rm lim}})^{-6}\Gamma_3^2
(\frac{\varepsilon_h}{70 {\rm GeV}})^{-1}
(\frac{\varepsilon_p}{\rm 2 MeV})^{-1}
\end{array}
\end{equation}
where $\tau_{\gamma\gamma}\simeq ({\Gamma}/{\Gamma_{\rm
lim}})^{-6}$ has been used in the last step and $\Gamma_{\rm lim}$
is the minimum bulk Lorentz factor of the outflow required by the
transparency of the high energy photon of energy $\varepsilon_h$.
As $\tau_{\gamma\gamma}(\varepsilon_h=70 {\rm GeV})\la 1$, we
obtain an upper limit of the $p\gamma$ efficiency, i.e. $f_\pi\la
2\times10^{-3}$. If the prompt MeV-10 GeV emission is interpreted
as arising from the secondary emission of hadronic process, one
would need an unreasonably large energy budget in relativistic
protons.

\section{The maximum synchrotron photon energy and its implications}
First we calculate the maximum synchrotron photon energy that can
be reached in GRB shocks. We assume that the relative motion
between {the upstream and downstream plasma} is only mildly
relativistic (such as in internal shocks).  {For an electron being
shock accelerated, the residence time in downstream and upstream
regions }are respectively, $t'_{d}=\kappa_d \varepsilon'_{e,d}/q
B_d c$ and $t'_{u}=\kappa_u \varepsilon'_{e,d}/q B_u c$, where
$\varepsilon'_e$ is the energy of accelerated electrons, $B_d$ and
$B_u$ are respectively the magnetic fields in the shock downstream
and upstream, and $\kappa_{d,u}\ga 1$ parameterizes the efficiency
of shock acceleration, with $\kappa_{d,u}\simeq 1$ corresponding
to the fastest shock acceleration-- the Bohm diffusive shock
acceleration with the scattering mean free path equal to the
particle gyroradius. It is generally assumed that the downstream
magnetic field is close to the equipartition with the shock
internal energy, while the value of upstream magnetic field is
less clear.  As $B_u\la B_d$, the total acceleration time is
dominated by upstream residence time, so $t'_{acc}\simeq\kappa
\varepsilon'_e/q B_u c$. The maximum energy of accelerated
electrons in each region is determined by equating the residence
time with the shorter one of the cooling time and the available
dynamic time, i.e. $t'_{acc}= \min\{t'_{cool}, t'_{dyn}\}$. The
cooling time in the downstream and upstream are, respectively,
$t'_{cool,d,u}={3m_e c}/({4\sigma_T(U_{B_{d,u}}+k(\gamma)
U_{\gamma})})$, where $U_{B_d}$ and $U_{B_u}$ represent the
magnetic field energy density in downstream and upstream
respectively, and $U_{\gamma}$ is the photon energy density. {In
downstream region}, $U_{B_d}\gg k(\gamma_{M,d}) U_{\gamma}$, so
the maximum electron energy is $\gamma_{M,d}=(\frac{6\pi
q}{\kappa_d \sigma_T B_d})^{1/2}$. {In upstream region}, the
magnetic field energy density could be lower than
$k(\gamma_{M,u})U_{\gamma}$, and in this case, the maximum
electron energy is $\gamma_{M,u}=(\frac{3 q B_u}{4\kappa_u\sigma_T
k(\gamma_{M,u}) U_{\gamma})})^{1/2}$, where
$k(\gamma_{M,u})=({\gamma_{\rm T}}/{\gamma_{M,u}})^{1/2}$. So
$\gamma_{M,u}=(\frac{3 q B_u}{4\kappa_u\sigma_T
U_{\gamma}\gamma_{\rm T}^{1/2}})^{2/3}$. As $B_d\ga B_u$, the
electrons radiate more efficiently in the downstream and therefore
the relevant maximum Lorentz factor with the observed radiation is
$\gamma_{M}=\min[\gamma_{M,u}, \gamma_{M,d}]$. Depending {on}
which of $\gamma_{M,u}$ and $\gamma_{M,d}$ is larger, we divide
the discussion into two cases:

i)The $\gamma_{M,d}\la\gamma_{M,u}$ case. The maximum synchrotron
photon energy is
\begin{equation}
h\nu_{syn, M}=0.2294\frac{3qB_d}{4\pi m_e c}\gamma_{M,d}^2\Gamma
=55 \left(\frac{1}{\kappa_d}\right) \Gamma_3 {\rm GeV},
\end{equation}
which is only dependent of the bulk Lorentz factor $\Gamma$ of the
relativistic outflow (0.2294 is the coefficient quoted from Wijers
\& Galama 1999). If the observed highest energy photon is produced
by synchrotron radiation, from $h\nu_{syn, M}\ga \varepsilon_h$,
we obtain
\begin{equation}
\kappa_d\la 0.8\Gamma_3 \left(\frac{\varepsilon_h}{70 {\rm
GeV}}\right)^{-1},
\end{equation}
which favors the Bohm diffusive acceleration if the bulk Lorentz
factor $\Gamma_3\la {a  few}$.

From the precondition, $\gamma_{M,d}\la\gamma_{M,u}$, we obtain a
lower limit of the upstream magnetic field in this case,
\begin{equation}
\begin{array}{ll}
B_u\ga \frac{4\kappa_u \sigma_T U_{\gamma}\gamma_{\rm
T}^{1/2}}{3q}\left(\frac{6\pi q}{\kappa_d \sigma_T
B_d}\right)^{3/4}
\\=500
(\frac{\epsilon_e}{\epsilon_B})^{3/8}\kappa_u\kappa_d^{-3/4}L_{\gamma,54}^{5/8}\Gamma_3^{-13/4}(\varepsilon_p/2
{\rm MeV})^{-1/2} t_{v,-2}^{-5/4} \, {\rm G}
\end{array}
\end{equation}

ii)For the case $\gamma_{M,u}\la \gamma_{M,d}$, the maximum
synchrotron photon energy is $h\nu_{syn,
M}(\gamma_{M,u})=0.2294\frac{3qhB_d}{4\pi m_e
c}\gamma_{M,u}^2\Gamma$. So from $h\nu_{syn, M}(\gamma_{M,u})\ga
\varepsilon_h$, we obtain a lower limit of the upstream magnetic
field,
\begin{equation}
\begin{array}{ll}
B_u\ga \left(\frac{4\pi m_e cE_h }{0.2294 \times 3 q B_d
\Gamma}\right)^{3/4}\frac{4\kappa_u \sigma_T U_\gamma \gamma_{\rm
T}^{1/2}}{3 q}\\\simeq 600 \kappa_u
(\frac{\epsilon_e}{\epsilon_B})^{3/8}\left(\frac{\varepsilon_h}{70
{\rm
GeV}}\right)^{3/4}L_{\gamma,54}^{5/8}\Gamma_3^{-4}(\frac{\varepsilon_p}{2
{\rm MeV}})^{-1/2} t_{v,-2}^{-5/4} {\rm G}.
\end{array}
\end{equation}
Combining this lower limit with the precondition $\gamma_{M,u}\la
\gamma_{M,d}$, we find that Eq.(15) is applicable only when $
\varepsilon_h\la 55 \kappa_d^{-1}\Gamma_3 {\rm GeV}$. In both
cases, the GRB shells that produce the prompt emission must have a
pre-shock magnetic field greater than $\sim 500 {\rm G}$ at a
radius of $R\sim 3\times 10^{14}{\rm cm} \Gamma_3^2 t_{v,-2}$. If
the field lines in the expanding shell are frozen and the width of
the shell is constant, the components then vary with distance as
$B_r\propto r^{-1}$ and $B_\theta\sim B_\phi\sim r^{-2}$. For an
initial magnetic filed of $B_0\sim 10^{15}{\rm G}$ within a volume
of radius of $10^6-10^7{\rm cm}$, the above limit is larger than
the $B_r$ component, but still within the $B_\theta$ or $B_\phi$
component. Of course, the above limit is also consistent with the
hypothesis that the upstream magnetic field is significantly
amplified by the particle streaming instability (Bell
2004){\footnote{Recently, Li \& Waxman (2006) constrained the
pre-shock magnetic fields of GRB afterglow shocks by synchrotron
X-ray afterglows,  which also implies that the pre-shock magnetic
fields may be amplified. }}. Interestingly, the shock compressed
upstream magnetic field, $B\sim 4 \Gamma_{sh} B_u\ga 1500 {\rm
G}$, is similar to the assumed equipartition magnetic field in
downstream (i.e. Eq.1), which means that the field compression due
to the shock is enough to explain the downstream magnetic field.

\section{Summary and Discussions}
The single-component  spectrum of GRB080916C from MeV to GeV puts
useful constraints on the emission mechanism. We found that the
synchrotron  mechanism from relativistic electrons is consistent
with the observed spectrum, while the simple one-zone electron IC
and hadronic models are less viable. In the synchrotron
interpretation, the SSC emission is found to be in the KN
scattering regime and as a consequence, the IC component is not
visible at high energies even though the magnetic field energy
density is smaller than that in the relativist electrons, i.e.
$\epsilon_B<\epsilon_e$, as obtained in our case. We also suggest
a scenario in which such a KN IC emission dominated regime can
explain the low energy photon spectral index of GRB 080916C.

The delay of high-energy gamma-ray emission relative to the
low-energy emission in GRB080916C is still a mystery in the
electron synchrotron scenario. It could be due to that the energy
distribution slope $p$ of electrons during the first time interval
(time a) is rather steep so that the high-energy emission is
suppressed or that the emission region has not become transparent
for high-energy gamma-rays at early times.

 {\acknowledgments We thank the referee for the valuable comments
that improve the paper. This work is supported by the 973 program
under grants 2007CB815404 and 2009CB824800,  the NSFC under grants
10873009, 10843007 and 10473010, the Foundation for the Authors of
National Excellent Doctoral Dissertations of China, the Qing Lan
Project, NCET and NASA grant NNX 08AL40G.}

\end{document}